\documentclass[preprint,review,12pt]{elsart}

\usepackage{graphics}
\usepackage{graphicx}

\usepackage{amssymb}
\usepackage{natbib}
\usepackage{lineno}
\newcommand{\grl}{Geophysical Research Letters}
\journal{Journal of Atmospheric and Solar-Terrestrial Physics on Space Climate}

\begin{document}

\begin{frontmatter}



\title{Long-term solar activity influences on South American rivers}


\author[iafe]{Pablo J.D. Mauas}
\author[iafe]{Andrea P. Buccino} 
\address[iafe]{Instituto de Astronom\'\i a y F\'\i sica del Espacio
  (CONICET-UBA),  C.C. 67 Sucursal 28, 1428,Buenos Aires, Argentina}

\author[inta]{Eduardo Flamenco}
\address[inta]{Instituto Nacional de Tecnolog\'\i a Agropecuaria,
  Rivadavia 1439, 1033,  Buenos
  Aires, Argentina}

\begin{abstract}
River streamflows are  excellent climatic indicators since they
integrate \textbf{precipitation} over large areas. Here we follow up on our
previous  study of the influence of solar activity on the flow of the
Paran\'a River, in South America.
We find that the unusual minimum of solar activity in
recent years have a correlation on very low levels in the Paran\'a's
flow, and we report historical evidence of low water
levels during the Little Ice Age. We also study data for the
streamflow of three other rivers (Colorado, San Juan and Atuel), and
snow levels in the Andes. We obtained that, after eliminating the
secular trends and smoothing out the solar cycle, there is a strong 
positive correlation between the residuals of both the Sunspot Number
and the streamflows,  as we obtained for the Paran\'a. 
Both results put together imply that higher solar activity
 corresponds to larger \textbf{precipitation}, both in summer and in
wintertime, not only in the large basin of the Paran\'a, but also in
the Andean region north of the limit with Patagonia.
\end{abstract}

\begin{keyword}
South American rivers, solar activity, streamflow.



\end{keyword}

\end{frontmatter}

\linenumbers

\section{Introduction}\label{sec.intro}
Usually, studies focusing on the influence of solar activity on
climate have concentrated on Northern Hemisphere temperature or sea
surface temperature. However, climate is a very
complex system, involving many other important
variables. Recently, several studies have focused in a different
aspect of climate: atmospheric moisture and related quantities like,
for example, \textbf{precipitation}.

Perhaps the most studied example is the Asian monsoon, where
correlations between solar activity and \textbf{precipitation} have been found
in several time scales. For example, \cite{2001Natur.411..290N} found
strong coherence between solar variability and the monsoon in Oman
between 9 and 6 kyr ago. \cite{2002E&PSL.198..521A} found that Indian
monsoon intensity followed the solar irradiance variability on
centennial time scales during the last
millennium. \cite{2003Sci...300.1737F} studied Holocene forcing of the
Indian monsoon, and found that intervals of weak (strong) solar
activity correlates with periods of low (high) monsoon \textbf{precipitation}.
On shorter time scales, \cite{1997GeoRL..24..159M}, found that, at
decadal-multi\-de\-ca\-dal time scales, the correlation between the El Ni\~no 3
index and the monsoon rainfall is stronger when solar irradiance is
above normal and {\bf{viceversa}}. Correlations between solar activity and
Indian monsoon in decadal time scales were also found by
\cite{2005GeoRL..3205813B} and \cite{2004GeoRL..3124209K}, among others.

\cite{2005Sci...308..854W} studied the monsoon in southern China over
the past 9000 years, and found that higher solar irradiance
corresponds to stronger monsoon. They proposed that the monsoon
responds almost immediately to solar changes by rapid atmospheric
responses to solar forcing.

All these studies reported a positive correlation, where periods of
higher solar activity correspond to periods of larger
\textbf{precipitation}. In contrast, \cite{2001E&PSL.185..111H} studied a
6000-year record of drought and \textbf{precipitation} in northeastern China,
and found that most of the dry periods agree well with stronger
solar activity and {\bf{viceversa}}. In the American continent, droughts in
the Yucatan Peninsula have been 
associated with periods of high solar activity and have even been
proposed to explain the Mayan decline \citep{2001E&PSL.192..109H}.

In the same sense, studies based on the water level of Lakes
Naivasha \citep{2000Natur.403..410V} and Victoria \citep{stager05} in
East Africa, report severe droughts during phases of high solar
activity and increased \textbf{precipitation} during periods of low solar
irradiation. To explain these differences it has been proposed that
increased solar irradiation causes more evaporation in equatorial
regions, enhancing the net transport of moisture flux to the Indian
sub-continent via monsoon winds \citep{2002E&PSL.198..521A}.

However, these relationships seem to have reversed sign around 200
years ago, as severe droughts developed over much of tropical Africa
during the Dalton sunspot minimum, ca. AD 1800-1820 \citep{stager05}.
Furthermore, \cite{2007JGRD..11215106S}
studied recent water levels in Lake Victoria, and found
that peaks in the $\sim$11-year sunspot cycle were accompanied by water
level maxima throughout the 20th century, due to the occurrence of
positive rainfall anomalies $\sim$1 year before solar maxima. Similar
patterns also occurred in at least five other East African lakes,
indicating that these sunspot-rainfall relationships were broadly
regional in scale.
 
A different approach was taken by \cite{2005MmSAI..76.1002M} who
proposed to study the streamflow of a large river, the
Paran\'a in southern South America, as an indicator of
\textbf{precipitation}.  In fact, flows of continental-scale rivers are excellent
climatic indicators since they integrate \textbf{precipitation}, infiltrations
and evapotranspiration over large areas and smooth out local
variations.  Signals of solar activity have recently been found with
spectral analysis techniques in the river Nile by \cite{Ruzmaikin2006},
who found a low-frequency 88-year variation present in solar
variability and in the Nile records. Similarly, \cite{2008JGRD..11312102Z}
found that the \textbf{discharge} of the Po river appear to be correlated with
variations in solar activity, on decadal time scales.

In \cite{2008PhRvL.101p8501M} (hereinafter Paper I) we presented the
results of our study of the Paran\'a. We found that the 
streamflow variability of the Paran\'a river has three temporal 
components: on the secular scale, it is probably part of the global climatic 
change, which at least in this region of the world is related with more humid 
conditions; on the multidecadal time scale, we found a strong correlation with 
solar activity, as expressed by the Sunspot Number, and therefore probably with 
solar irradiance, with higher activity coincident with larger discharges; on 
the yearly time-scale, the dominant correlation is with El Ni\~no. 

In the present paper we follow up on the study of the influence of
solar activity on the flow of South American rivers. In Section 2 we
expand in time the study of the multi-decadal component of the
Paran\'a's streamflow, to include the most recent years, which have shown
particularly low levels of solar activity. In Section 3 we study
other South American rivers, to see whether the influence extends to
other areas of the continent. Finally, in Section 4 we discuss the
implications of our findings.


\section{The multidecadal component of the Paran\'a's streamflow}\label{subsec.parana}

The Paran\'a is the fourth river of the world  according to streamflow (20 600
m$^3$/s), and the fifth  according to drainage area (3 100 000
km$^2$), which is the second largest in South America.
Its origin is in the southernmost part of the 
Amazon forest, from where it flows south collecting water from  territories in
Brazil, Bolivia, Paraguay and Argentina. Its outlet is in the
Plata River, a few kilometers north of the City of Buenos Aires. It
flows through heavily populated areas and it is navigated by overseas trade ships,
unlike other rivers of similar size. For these reasons,
its streamflow has been measured continuously during the last century.


As in Paper I, we analyze the streamflow data measured daily since
1904, at a gauging station located in the city of Corrientes, 900 km
north of the outlet of the Paran\'a. Since the Paran\'a's hydrological
year goes from September to August, with maximum streamflow in the
Southern Hemisphere's summer months of January, February and March,
our yearly series integrates the flow from September to August of the
next year.

In Paper I we found that in intermediate scales of decades, there is
a strong correlation between the Paran\'a's streamflow and solar activity, as 
expressed by the Sunspot Number ($S_N$)\footnote{Available at
\textsf{ftp://ftp.ngdc.noaa.gov/STP/SOLAR$\_$DATA/SUNSPOT$\_$NUMBERS}.},
with larger solar activity corresponding to larger streamflow. We found a similarly strong
correlation with the irradiance reconstruction by \citet{2005ApJ...625..522W}.

\begin{figure}[htb!]
\resizebox{\hsize}{!}{ \includegraphics{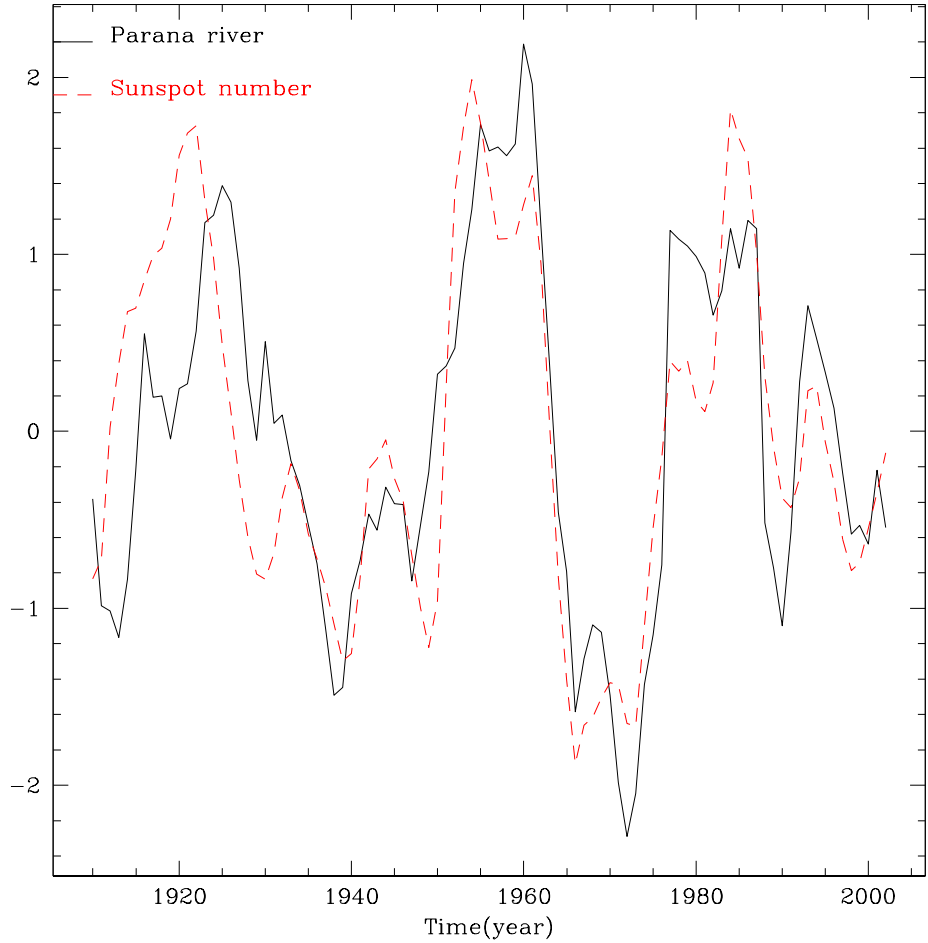}}
 \caption{ The detrended time series for the Paran\'a's streamflow
 (full line) and the Sunspot Number (dashed line). The detrended
 series were obtained by subtracting from each data series the
 corresponding secular trend and were smoothed by an 11-yr-running
 mean to eliminate the solar cycle. Both series were standardized by
 subtracting the mean and dividing by the standard deviation, to avoid
 introducing arbitrary free parameters.  The Pearson's correlation
 coefficient is R=0.78. }\label{parana_multi}
\end{figure}

 We further explore this correlation in  Fig. \ref{parana_multi},
which is an update of Fig. 2 of Paper I, including 4 more years of
data. To retain only the intermediate scale, we first computed the 
secular trends with a low-pass Fourier filter with a 50 years
cut-off, as shown in Fig. 1 of Paper I, which was substracted from
the data. Then, we performed an 11-year-running mean to smooth out
the solar cycle (for this reason, only data for the period 1909-2003 are shown).
In this way, both high and low frequencies have been filtered out of
the data in Fig. \ref{parana_multi}, which only retain the
variations in timescales between 11 and 50 years.

When plotting together different quantities, the offset and the
relative scales are free parameters which are usually arbitrarily
introduced. To avoid these two artificial parameters, as a final step
we have standardized the quantities by subtracting the mean and
dividing by the standard deviation of each series shown, for the whole
period 1909-2003. More details can be found in Paper I.

The correlation between the Paran\'a's streamflow and the Sunspot
Number in Fig. \ref{parana_multi} is quite remarkable. In
fact, the correlation coefficient between both series is R=0.78,
significant to a 99\% level. 

 We point out that this correlation is found in the intermediate time
scale. On longer timescales, both the Paran\'a's discharge and solar
activity are larger in the last decades than in the first ones of the
20th century, and these increases are not correlated (for a discussion,
see Paper I). On the yearly timescale, the dominant factor influencing
streamflow's variations is El Nin\~no (again, see Paper I for
details). The results shown in  Fig. \ref{parana_multi} show that {\it
decades} of larger discharge correspond to  {\it decades} of higher
activity, with these variations overimposed on the corresponding
secular trends.

It can be seen that the correlation is
still found in the most recent years. In particular, in the period
1995-2003 both the mean Sunspot Number and the Paran\'a's streamflow
have decreased by similar amounts. In fact, Solar Cycle 23 was the
weakest since the 1970s, and the onset of Solar Cycle 24 was delayed
by a minimum with the largest number of spotless days since the
1910s.  Morover, $S_N$ for the years 2008 and 2009 (2.9 and 3.1,
respectively), have been the lowest since 1913. Similarly, the mean
levels of the Paran\'a were also the lowest since the 1970s. 

We have also tested the correlations between the Paran\'a's discharge and
the neutron count at Climax, Colorado\footnote{Available at 
\textsf{http://www.env.sci.ibaraki.ac.jp/ftp/pub/WDCCR/STATIONS/climax}.}, which is a
direct measure of the galactic cosmic rays (GCR) flux. Of course,
since neutron count is correlated with sunspot numbers, we found a
correlation with Paran\'a's streamflow. However, the correlation with
$S_N$ is larger, pointing to a more direct correlation with solar
irradiance than with GCR.

The relationship between smaller solar activity and low Paran\'a's
discharge can also be found in historical records. For example, low
discharges were reported during the period known as the Little Ice Age
(LIA). In particular, a traveler of that period recalls in his diary 
that in the year 1752 the streamflow was so small that
the river could not even be navigated by the ships of that time, which
were less than 5 ft draft, to be compared with ships up to 18 ft draft
that can navigate it at present as far north as Asunci\'on in Paraguay
\citep{Iri99}. The fact that the LIA coincided with reduced
\textbf{precipitation} in this region has been found in different climatic
records (e.g. \citealt{piovano09} and references therein).
It is well known that the LIA coincided with, and perhaps was
caused by, low solar activity \citep{1976Sci...192.1189E}.

\section{The Colorado river basin}

\begin{figure}[htb!]
\centering
\resizebox{\hsize}{!}{\includegraphics{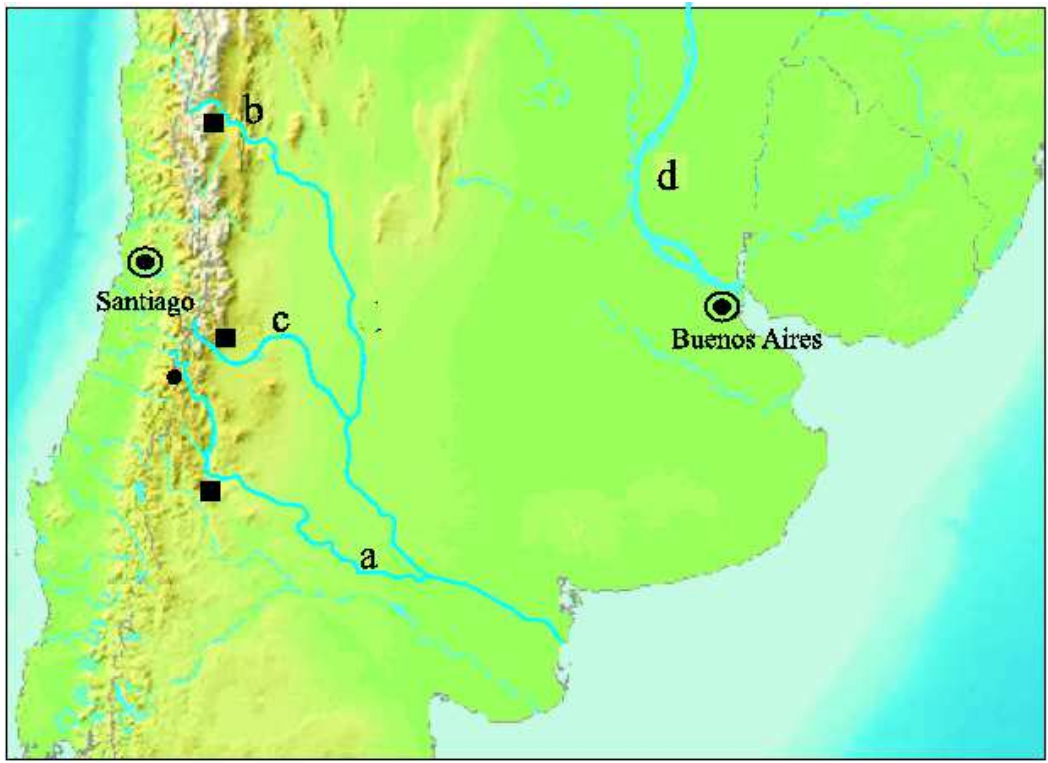}}
\caption{Colorado hydrologic system. The rivers under study are marked
  in the figure: a. Colorado, b. San Juan, c. Atuel and
  d. the lower part of the Paran\'a river. The stream flow ($\blacksquare$) and snow
  ($\bullet$) measuring stations are also indicated.}\label{cuencacol}
\end{figure}

Here we study the streamflow of the Colorado river, and two of its
tributaries, the San Juan and the Atuel rivers (see
Fig. \ref{cuencacol}). We also analyze snow 
levels, measured near the sources of the Colorado.

 The Colorado river marks the north boundary of the Argentine
Patagonia, separating it from the Pampas, to the northeast, and the
Andean region of Cuyo, to the Northwest.  Its origin is on the eastern
slopes of the Andes Mountains, from where it flows southeast until it
discharges in the Atlantic Ocean.
The Atuel, 
which originates in the glacial Atuel Lake, at 3250 m above sea level in
the Andes range, 
and the 500 km long San Juan river, 
join the Colorado downstream of it's gauging station. Therefore, 
the data given by the three series are not directly related. 

In Table \ref{tab.rios} we list the mean stream flow and
  drainage basin area of the Colorado, Atuel and San Juan rivers. We also
  include the time interval of the stream flow records plotted in
  Fig. \ref{riosnie} and the geographical coordinates
  of the gauging stations. 

\begin{table}
\centering
\begin{tabular}{|c c c c c c|}
\hline
\hline
River      & $\langle S \rangle$& $A$ & $T$&  \multicolumn{2}{c|}{
  Gauging Station} \\
      & (m$^3$/s) &  (km$^2$)   &       &    &       \\
\hline 
Colorado  & 150 & 15 300  & 1940-2006 & Buta Ranquil  & 37$^\circ$
06' S  69$^\circ$ 44' W\\
Atuel   & 32 & 3800 &   1916-1999 & La Angostura & 35$^\circ$
02' S 68$^\circ$ 52' W\\
San Juan & 56 & 25700 & 1909-2005 & KM 47,3 & 31$^\circ$
32' S 68$^\circ$ 53' W\\
\hline
\hline
\end{tabular}
\caption{$\langle S \rangle$: Mean streamflow.  $A$:drainage basin area. $T$:
  time interval f stream flow records.}\label{tab.rios}
\end{table}

\begin{figure}[htb!]
\centering
\resizebox{\hsize}{!}{\includegraphics{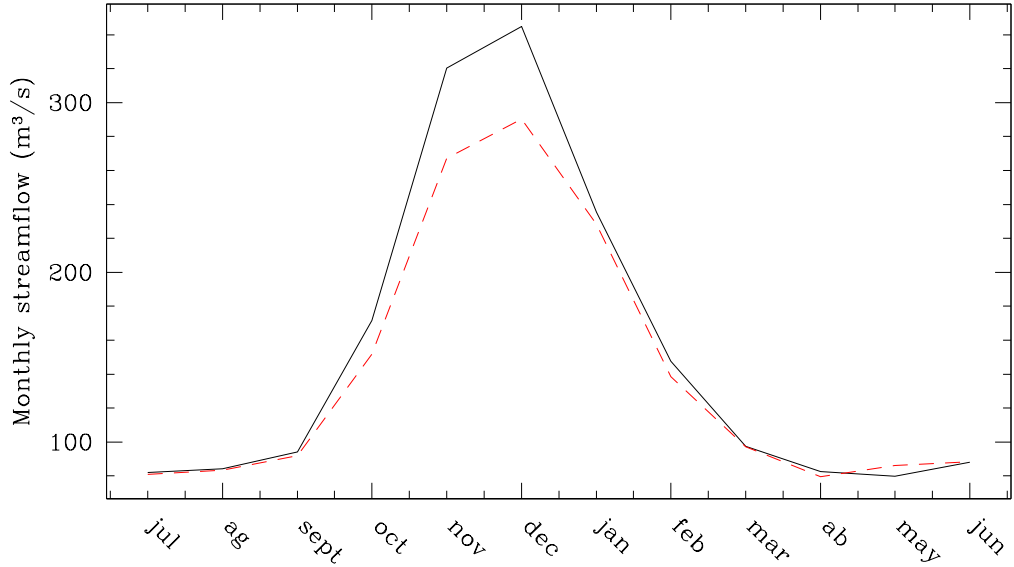}}
\caption{Monthly Colorado's mean streamflow. The solid (dashed) curve
  was obtained considering only the years when the multidecadal
  component of $S_N$ shown in Fig. \ref{parana_multi} is high
  (low), \it {i.e.} larger (smaller) than 0.5
$\sigma$ above (below) the mean value.}\label{colo_mes}
\end{figure}

\begin{figure}[htb!]
\centering
\includegraphics[width=1.\textwidth]{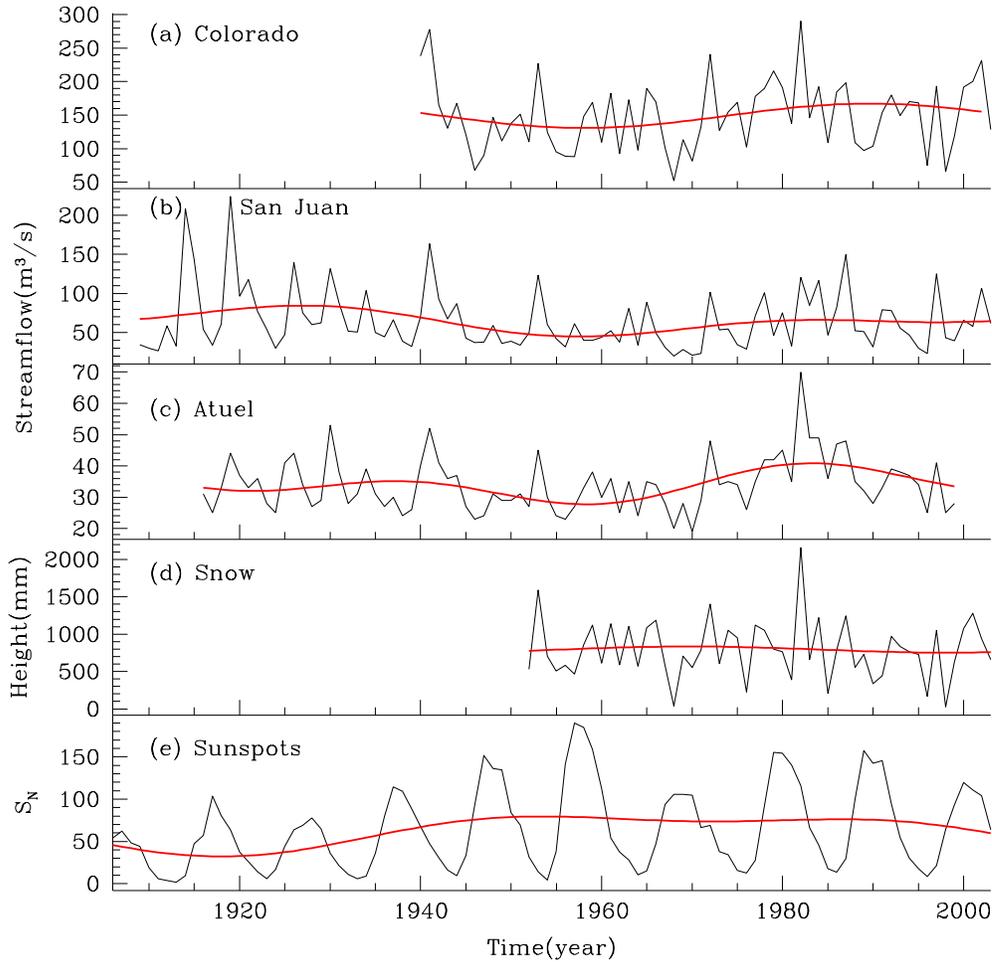}
\caption{Time series of the Colorado, San Juan and Atuel's
streamflow, the
snow height measured at Valle Hermoso and  the Sunspot Number. The
long-term trend of each series is marked with a heavy
line. }\label{riosnie}
\end{figure}

Unlike the Paran\'a, whose streamflow is directly related to
\textbf{precipitation}, the regime of all these rivers is dominated by
snow melting, and their streamflows reflect \textbf{precipitation}
accumulated during the winter, and melted during spring and
summer. For this reason, the streamflows are largest during summer,
and the hydrological year for these rivers goes from July to June next
year. This can be seen In \mbox{Fig. \ref{colo_mes}}, \textbf{where}
we show the mean monthly \textbf{flow} of the Colorado. In the figure
we separately plot the flow for years when the multidecadal component
of the sunspot number shown in Fig. \ref{parana_multi} is high (low),
i.e. larger (smaller) than 0.5 $\sigma$ above (below) the mean value.
It can be seen that during the \textbf{decades} with larger activity,
the streamflow is larger from September to December, when most of the
melting takes place, and remains almost constant during the rest of
the year.

To directly study the snow \textbf{precipitation}, we complete our data with
measurements of the height of snow accumulated at Valle Hermoso (35$^\circ$
15' S; 70$^\circ$ 20' W), in
the Andes at 2250 m above Sea level, close to the origin of the
Colorado (see Fig. \ref{cuencacol}), which were measured in situ at
the end of the winter since 1952. 
 In fact, the correlation between the streamflow of the Colorado and
the snow height is very good, with a correlation coefficient R=0.87,
significant to a 99\% level. Correlation coefficients between the snow
data and the Atuel and San Juan streamflows are R=0.76 and R=0.64,
respectively. Since Valle Hermoso  is placed closer to the origin of
the Colorado, and closer to the Atuel than to the San Juan, this
progressive reduction of the correlation is to be expected.

In Fig. \ref{riosnie} we plot the yearly time series of the streamflow of
the Colorado, San Juan and Atuel rivers, the snow height and the Sunspot
number. We also
show the variation in the longest scales, obtained with a low-pass
Fourier filter, as we did for the
Paran\'a.  However, since the length of the time series is not the same
for every set of observations, we could not apply a uniform filter for
all of them. In all cases, the cut-off was taken as half the length of
the observations (33 years for the Colorado, 40
years for the Atuel, 50 years for the San Juan and the Sunspot Number and  28 years for the snow)

\begin{figure}[htb!]
\includegraphics[width=.6\textwidth]{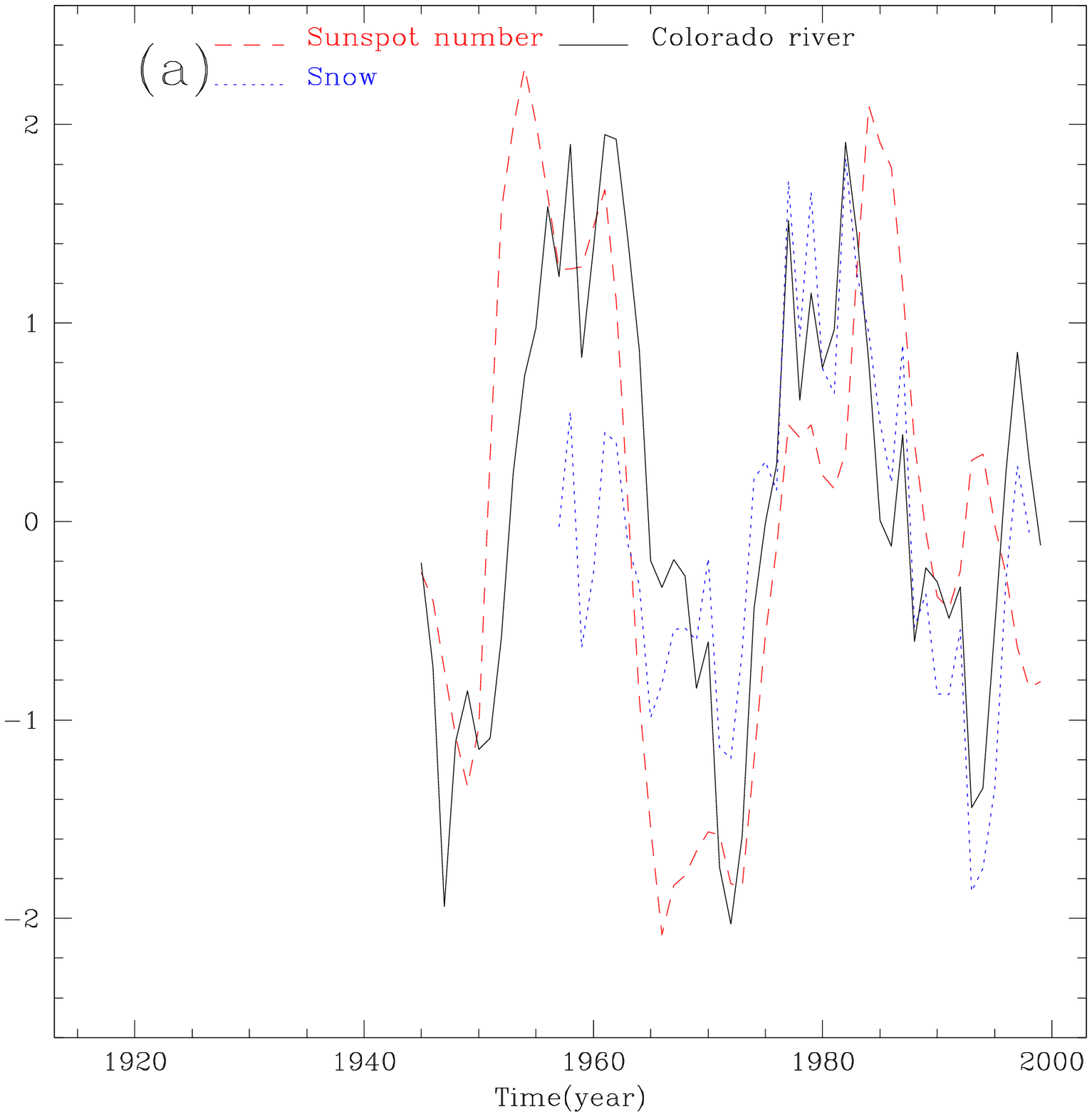}
\includegraphics[width=.6\textwidth]{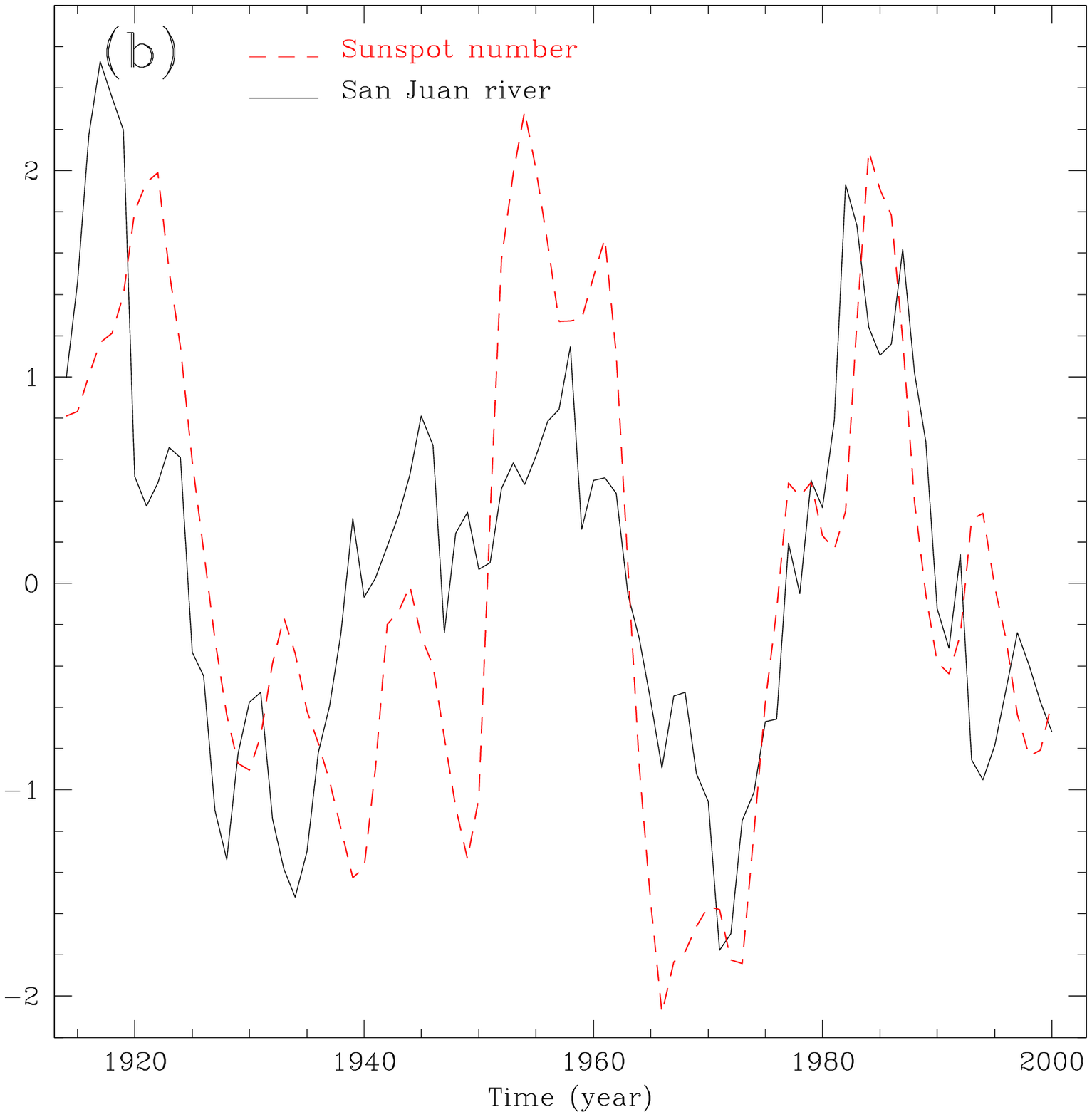}
\includegraphics[width=.6\textwidth]{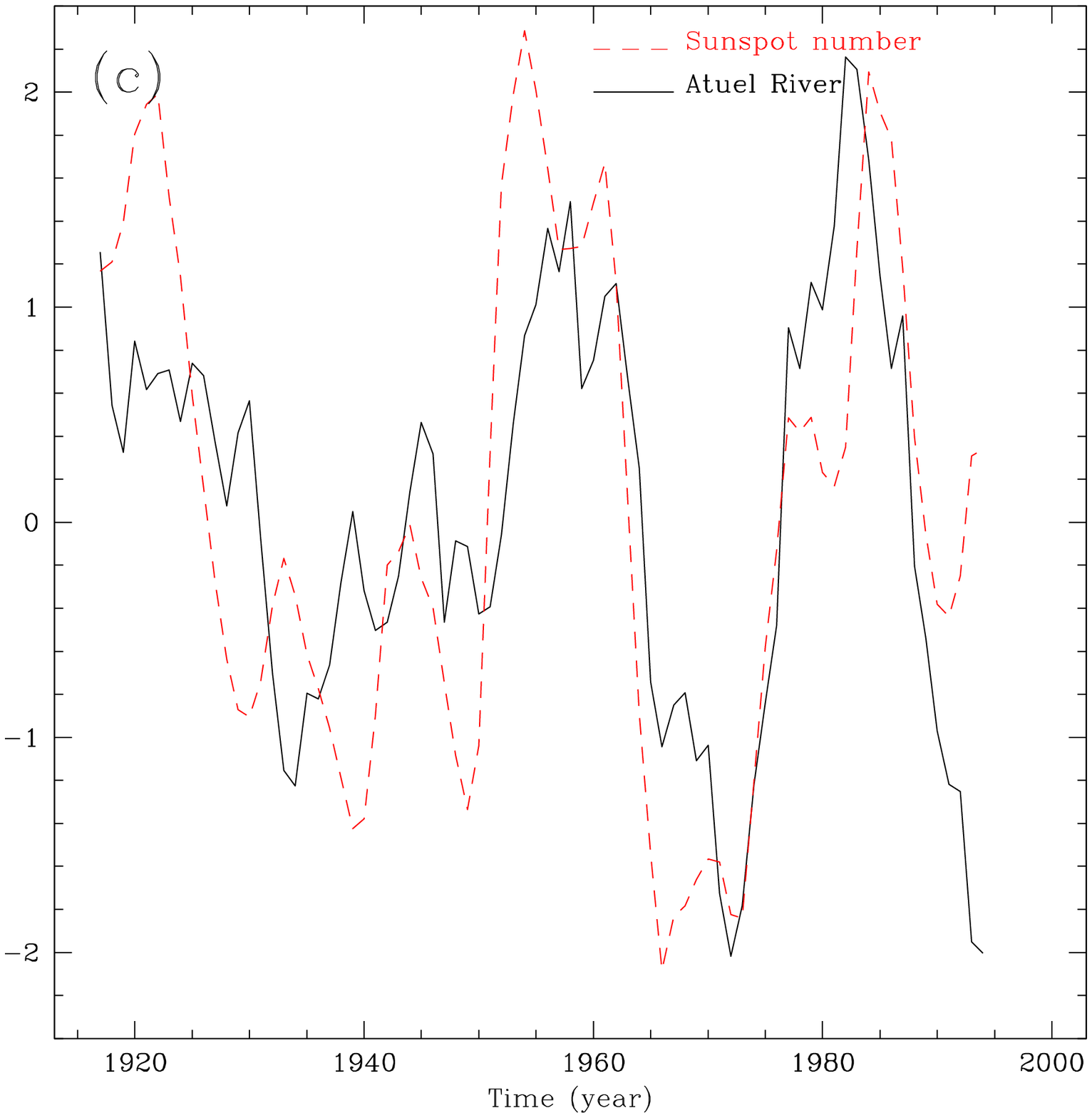}\hfill
\begin{minipage}{7cm}{\vspace{-5cm}\caption{ The detrended  streamflows (full lines) compared with Sunspot Number
   (dashed lines).  In panel (a) the snow level is also shown (dotted
   line). The data were smoothed with an 11-year running mean,
   detrended by substracting the long term component. All
   series were standardized by subtracting the mean and dividing by
   the standard deviation of each series shown for the period
   1971-2000.}\label{todo_multi}}\end{minipage}
\end{figure}

In \mbox{Fig. \ref{todo_multi}} we compare the multidecadal component
of the streamflows with the corresponding series for the sunspot
number. In each case, we smoothed out the solar cycle with an 11-year
running mean, and we detrended the series by subtracting the long term
component shown in Fig. \ref{riosnie}. Finally, we standardized the
data by subtracting the mean and dividing by the standard deviation of
each series shown for the period 1971-2000, suggested by the
\emph{World Meteorological Organization} as standard reference.  In
the panel corresponding to the Colorado, we also include the snow
height.

It can be seen that in all cases the agreement is remarkable. The
correlation coefficients are 0.59, 0.47, 0.67 and 0.69 for the
Colorado, the snow level, the San Juan and the Atuel, respectively, all
significant to the 96-97\% level. Therefore, also in these cases we
found a relation between solar activity, on one hand, and the
streamflow of these rivers and snow level, on the other, as we found
for the Paran\'a.

Probably, the {\bf most important correlation is the one with snow
level, and the correlations with the rivers' streamflow are indirect
consequences of the variations in precipitations.}
We should point out that climate in this area is correlated with the
conditions over the \textbf{equatorial} Pacific, as measured by El
Ni\~no. This correlation was studied for the Diamante River, also a
tributary of the Colorado, by \cite{1999WRR....35.3803B}.

In particular, the peaks in the snow level and the streamflows
  (see Fig. \ref{riosnie})
in the year 1982/3 coincide with a very strong El Ni\~no event, 
which caused a huge flood in the Paran\'a's basin, as we discussed in
Paper I. Correlation coefficients between our data and el Ni\~no 1+2 index in November (at the beginning of the
austral Summer), are R=0.51, R=0.60, R=0.60, and R=0.61, for snow
level and the Colorado, Atuel and San Juan rivers, respectively. 

Although all these rivers have maximum streamflow during Summer,
there is a big difference, however, between the regimes of the Paran\'a 
and the remaining rivers: for these ones, the important factor is the
intensity of the \textbf{precipitation} occurring in the winter months, from
June to August. For the Paran\'a, what is most important is the level
of the \textbf{precipitation} during the summer months. It is also worth
noticing the sense of the relationship: here again, stronger activity
coincides with larger \textbf{precipitation}.

\section{Discussion}

In this paper we analyzed the influence of solar activity in the
streamflow of South American rivers of different regimes. 
First, we extended in time the study of the correlation between
Sunspot Number and the Paran\'a's streamflow we reported in Paper I.
On one hand, we found that the unusual minimum of solar activity in
recent years have a correlation on very low levels in the Paran\'a's
flow. On the other, we reported historical evidence of low water
levels during the Little Ice Age. 
We also found that the correlation is stronger with sunspot number
than with neutron count, which confirms that what is affecting climate
is most probably solar irradiance, and not GCRs.

The fact that the river's behaviour follows $S_N$ through one more
minimum strongly enhances the significance of the correlation and its
predictive value. In particular, the low levels of activity expected for Solar
Cycle 24 anticipate that the dry period in the Paran\'a will continue
well into the next decade.

To study whether the solar influence extends to other areas
of the continent, we analyzed the streamflow of three South American
rivers: the Colorado  and two of its tributaries, the
San Juan and Atuel rivers. We also used snow level from a station at
the origin of the Colorado.  We obtained that, after eliminating the
secular trends and smoothing out the solar cycle, there is a strong 
correlation between the residuals of both the Sunspot Number and the
streamflows. In all cases, the correlation we found on multi-decadal time
scales is positive, i.e., higher solar activity corresponds to larger
snow accumulation and, therefore, to larger discharges of all these
rivers, as we obtained for the Paran\'a river. 

Therefore, both results put together imply that higher solar activity
 corresponds to larger \textbf{precipitation}, not only in the large 
basin of the Paran\'a, but also in the Andean region north of the
limit with Patagonia. Furthermore, since streamflow variability of
rivers on central Chile are controlled by the same mechanisms that
regulate the rivers studied in this paper, one might expect the same
correlation to be found west of the Andes.

 Solar activity can affect \textbf{precipitation} through the position of
the Inter Tropical Convergence Zone (ITCZ), which has been shown to
correlate with variations in solar insolation
\citep{2004GeoRL..3112214P,2001Sci...293.1304H}. In fact, it has been
proposed that a displacement southwards of the ITCZ would increase
\textbf{precipitation} in southern tropical South America
\citep{2006GeoRL..3319710N}. We point out that increased
\textbf{precipitation} occur both in the Southern Hemisphere's summer when the
ITCZ is over the equator, close to the origin of the Paran\'a, and in
wintertime, when the ITCZ displaces north, and \textbf{precipitation} increase
further South.

\ack{We thank the Subsecretar\'\i a de Recursos H\'\i dricos de la Naci\'on 
(Argentina), and particularly Lic. Daniel Cielak, for facilitating the 
data used in this study.}

\bibliographystyle{icarus}



\end{document}